\documentclass[twocolumn,showpacs,preprintnumbers,amsmath,amssymb]{revtex4}

\usepackage{graphicx}
\usepackage{dcolumn}
\usepackage{bm}
\usepackage{amssymb}
\usepackage{indentfirst}
\usepackage{psfig,color}
\usepackage{epsfig}
\usepackage{epsf}
\usepackage{graphicx}

\def\Vec#1{{\bf #1}}

\newcommand{\ssh}{\!\!\!/}

\preprint{USM-TH-175}

\begin{document}

\title{Generalized Distribution Amplitudes at the $Z$ pole}
\author{Zhun Lu} \author{Ivan
Schmidt}\affiliation{Departamento de F\'\i sica, Universidad
T\'ecnica Federico Santa Mar\'\i a, Casilla 110-V, Valpara\'\i so,
Chile}

\begin{abstract}
We investigate the exclusive two-pion production at the $Z$-pole
through the process $e^+e^-\rightarrow Z \rightarrow \pi\pi \gamma$.
In the kinematical region where the invariant mass of the two pions
is much smaller than the mass of the $Z$ boson, the process can be
factorized into the convolution of a hard coefficient and a soft
matrix element, the generalized distribution amplitude of two pions.
We calculate the differential cross-section of charged pion pair
production and neutral pion pair production and find the former to
be much larger than the later. We show that combining the
measurements of charged pion pair production and neutral pion pair
production at the $Z$-pole provides a convenient approach to access
the $C$-odd part of the $2\pi$ distribution amplitudes.
\end{abstract}


\maketitle

\section{Introduction}

The QCD description of exclusive processes has been successfully
extended to meson pair production through two-photon collisions
$\gamma^*\gamma\rightarrow M\bar{M}$, in the kinematical region of
large virtuality of one photon and of small center of mass
energy~\cite{Diehl98,muller,freund00}. The scattering amplitude of
the process can be factorized into a perturbative calculable hard
process $\gamma^*\gamma\rightarrow q \bar{q}$ or
$\gamma^*\gamma\rightarrow gg$, and non-perturbative matrix elements
describing the transition of the two partons into a hadron pair,
which are called generalized distribution amplitudes
(GDAs)~\cite{Diehl98}, to emphasize their close connection with the
distribution amplitudes introduced many years ago in the description
of exclusive hard process~\cite{bl80}. The GDAs are related by
crossing~\cite{teryaev} to the generalized parton distributions
(GPDs)~\cite{dm,muller,dvcs}, which enter the factorization of
deeply virtual Compton scattering and hard exclusive
processes~\cite{cfs97}, and encode information of the quark orbital
angular momentum contributing to the nucleon spin~\cite{ji97}. A
systematic investigation of the pions GDAs (2$\pi$DAs) has been
performed in Refs.~\cite{Diehl00, Polyakov99}, and the corresponding
phenomenology at $e^+e^-$ colliders has been
developed~\cite{Diehl00} in detail. GDAs can also be defined for
more complicated systems. The case of three pions has been studied
in~ \cite{pt00}, two baryon production in~\cite{dkv03} (although in
a different kinematical region), and the case of two vector mesons
($\rho$ mesons) $\gamma^*\gamma \rightarrow \rho\rho$
in~\cite{apt04}, which has been measured first by the L3
Collaboration~\cite{L3} at LEP and the data has been analyzed
in~\cite{apt04}.

GDAs appear in a variety of hard scattering processes other than the
two-photon process. Studies have also been performed on exotic meson
production~\cite{astw}, and exclusive heavy mesons decaying to
multimesons~\cite{cl04}, where a large scale is set by the heavy
quark mass. In this paper we consider another process that has not
been studied so far, the pion pair production at the $Z$-pole:
$e^+e^-\rightarrow Z\rightarrow\pi\pi\gamma$. In this process the
mass of the $Z$ boson provides a natural hard scale. Thus in the
kinematical region where the invariant mass of pion pair $W$ is
small, the process can be separated into a hard subprocess
$Z\rightarrow q\bar{q}\gamma$ and a non-perturbative two-pion
distribution amplitude, analogous to $\gamma^*\gamma\rightarrow
\pi\pi$. We give a leading order calculation for neutral and charged
pion pair production at the $Z$-pole. Using an available model for
the 2$\pi$DAs we find that the cross-section for charged pion pair
production is much larger than the case of neutral pion pair, which
means that the process is dominated by the isovector channel, that
is, the contribution from the $C$-odd part of the $2\pi$DAs. The
same one also contributes to the hard exclusive electroproduction of
pion pairs~\cite{Polyakov99,lppsk}. Our study shows that combining
the measurement of charged pair and neutral pair production at the
$Z$-pole can give information of the $C$-odd part of the $2\pi$DAs.
The Bremsstrahlung process $e^+e^-\rightarrow
\gamma^*\gamma\rightarrow\pi^+\pi^-\gamma$ is also considered in our
analysis.

\section{Analysis of the process $Z\rightarrow\pi\pi\gamma$}

\begin{figure}

\begin{center}
\scalebox{0.73}{\hspace{-0.5cm}\includegraphics{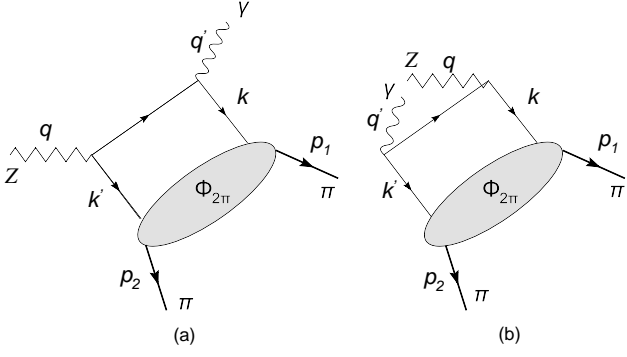}}
\caption{\small Factorization of pion pair production at the
$Z$-pole} \label{zpole}
\end{center}

\end{figure}

\begin{figure}

\begin{center}
\scalebox{0.95}{\includegraphics{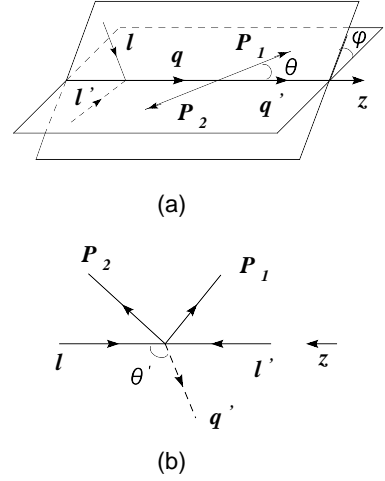}} \caption{\small The
kinematics of pion pair production. (a) the c.m. frame of the pion
pair, (b) the c.m. frame of $e^+e^-$. } \label{kine}
\end{center}

\end{figure}

The underlying process we study is:
\begin{equation}
Z(q)\rightarrow \pi(p_1)+\pi(p_2)+ \gamma(q^\prime).
\end{equation}
The photon emission allows for a small invariance mass to be
transferred to the pion pair. Denoting by $q$ and $q^\prime$ the
momenta of the $Z$ boson and final state photon, respectively, we
use $p_1$ and $p_2$ to denote the momenta of the two-pion mesons,
and define the sum of the two-pion momenta as $P$. The lowest order
diagrams of pion pair production at the $Z$-pole are shown in
Fig.~\ref{zpole}. We choose the pion pair center of mass frame
(shown in Fig.~\ref{kine}a) as the reference frame, and define the
$Z$ boson momentum direction as the $z$ axis. In this frame we have
\begin{eqnarray}
q&=&\frac{Q}{\sqrt{2}}v+\frac{Q}{\sqrt{2}}v^\prime,~~~
q^\prime=\frac{Q^2-W^2}{\sqrt{2}Q}v^\prime,\nonumber \\
P&=&\frac{Q}{\sqrt{2}}v+\frac{W^2}{\sqrt{2}Q}v^\prime,
\nonumber \\
p_1^\mu&=&\frac{\zeta Q}{\sqrt{2}}v^\mu+\frac{(1-\zeta)W^2}{\sqrt{2}Q}v^{\prime\mu}+\frac{\Delta_T^\mu}{2},\nonumber\\
p_2^\mu&=&\frac{(1-\zeta)Q}{\sqrt{2}}v^\mu+\frac{\zeta
W^2}{\sqrt{2}Q}v^{\prime\mu}-\frac{\Delta_T^\mu}{2},
\end{eqnarray}
respectively. Here $v$ and $v^\prime$ are two lightlike vectors
which satisfy: $v^2=v^{\prime 2}=0$, $v\cdot v^\prime=1$, $W$ is the
invariant mass of the pion pair and $q^2=Q^2$. The skewness
parameter $\zeta$ is the momentum fraction of plus momentum carried
by $\pi(p_1)$ with respect to the pion pair:
\begin{equation}
\zeta=\frac{p_1^+}{P^+}=\frac{1+\beta \cos
\theta}{2},~~~~\beta=\sqrt{1-\frac{4m_\pi^2}{W^2}}.
\end{equation}

The hadronic tensor of the process is
\begin{eqnarray}
iT^{\mu\nu}&=&-\int d^4xe^{-iq\cdot x}\langle
\pi(p_1)\pi(p_2)|TJ_{NC}^\mu(x)J_{EM}^\nu(0)|0\rangle
\nonumber\\
&=&\frac{P}{2\pi}^+\int dzH^{\mu\nu}_{\alpha\beta}(z,q,q^\prime)\int
dx^-e^{-izP^+x^-}\nonumber\\
&&\times\langle\pi(p_1)\pi(p_2)|\bar{\psi}_\beta(x^-)\psi_\alpha(0)|0\rangle,
\end{eqnarray}
which is expressed as a convolution of a hard coefficient function
and a soft correlation function, $z=k^+/P^+$ is the momentum
fraction of the quark with respect to the hadronic system.
$J_{NC}^\mu$ and $J_{EM}^\nu$ are the neutral current and
electromagnetic current respectively. The lignt-cone coordinates
used here are defined as $a^+=a \cdot v^\prime$, $a^-= a \cdot v$.
In the above equation the gauge-link ensuring the gauge invariance
of the definition is implied.

The hard coefficient function can be calculated from the subprocess
$Z\rightarrow q \bar{q} \gamma$ directly:
\begin{eqnarray}
H^{\mu\nu}&=&\frac{ie_qg}{2\cos \theta_W}\left
[V^\mu\frac{k\ssh-q\ssh}{(k-q)^2+i\epsilon}\gamma^\nu\right.\nonumber
\\&&\left.+
\gamma^\nu\frac{q\ssh-k^\prime\!\ssh}{(q-k^\prime)^2+i\epsilon}V^\mu\right
]\nonumber\\
&=&\frac{ie_qg}{2\sqrt{2}\cos \theta_W}\left [
V^\mu\frac{(z-1)Qv\ssh-Qv^\prime\!\ssh}{(1-z)Q^2}\gamma^\nu\right.\nonumber\\
&&\left.+\gamma^\nu\frac{zQv\ssh+Qv^\prime\!\ssh}{zQ^2}V^\mu\right
],
\end{eqnarray}
where $g = e/\sin\theta_W$ is the weak coupling constant,
$V^\mu=\gamma^\mu(c^q_V-\gamma_5c^q_A)$ is the $Z$-boson-quark
vertex, with $q$ denoting the quark flavor. The vector and
axial-vector coupling to the $Z$ boson are given by:
\begin{eqnarray}
c^q_V &=& T_3^q-2Q^q\sin^2\theta_W, \label{cqv}\\
c^q_A &=& T_3^q.\label{cqa}
\end{eqnarray}
Here $T_3^q$ =$+1/2$ for $q=u$, and $-1/2$ for $q=d$ and $s$, $Q^q$
is the electric charge of quark $q$ in units of the electron charge.

 Expanding the above
equation we arrive at:
\begin{eqnarray}
H^{\mu\nu}&=&\frac{ie_qg}{2\sqrt{2}\cos \theta_WQ}\left \{ \left
[c^q_VS^{\mu\nu\alpha\beta}\frac{1-2z}{z(1-z)}\right .\right.\nonumber\\
&& +\left.c^q_A\epsilon^{\mu\nu\alpha\beta}\frac{i}{z(1-z)}\right
]v^\prime_\alpha\gamma_\beta + 2 i
c^q_A\epsilon^{\mu\nu\alpha\beta}v_\alpha\gamma_\beta \nonumber\\&&-
2 i c^q_V \epsilon^{\mu\nu\alpha\beta}v_\alpha\gamma_\beta \gamma_5
+\left [c^q_A S^{\mu \nu \alpha \beta} \frac{2z-1}{z(1-z)}\right. \nonumber\\
&& \left.\left.- c^q_V \epsilon^{\mu \nu \alpha
\beta}\frac{i}{z(1-z)}\right ]v^\prime_\alpha\gamma_\beta
\gamma_5\right \},\label{hardcoef}
\end{eqnarray}
where we have used the identities:
\begin{eqnarray}
\gamma^\mu\gamma^\alpha\gamma^\nu&=&S^{\mu\alpha\nu\beta}\gamma_\beta+i\epsilon^{\mu\alpha\nu\beta}\gamma_5
\gamma_\beta,\nonumber\\
\gamma^\mu\gamma^\alpha\gamma^\nu\gamma_5&=&S^{\mu\alpha\nu\beta}\gamma_\beta\gamma_5-i\epsilon^{\mu\alpha\nu\beta}
\gamma_\beta,\nonumber\\
S^{\mu\alpha\nu\beta}&=&(g^{\mu\alpha}g^{\nu\beta}+g^{\nu\alpha}g^{\mu\beta}-g^{\mu\nu}g^{\alpha\beta}).
\end{eqnarray}
Convoluting $H^{\mu \nu}$ with the soft matrix element, and keeping
the leading twist contribution (the high-twist contributions are
irrelevant at the $Z$-pole), we get the hadronic tensor:
\begin{eqnarray}
&&T^{\mu \nu}=-\frac{g}{4 \cos\theta_W}\sum_q ee_q \left \{c^q_V
g_\perp^{\mu \nu} \int_{0}^{1} dz\frac{2z-1}{z(1-z)}\right
.\nonumber\\
& & \left .\times\Phi_q(z,\zeta,W^2) -ic^q_A\epsilon_\perp^{\mu \nu}
\int_{0}^{1} dz\frac{1}{z(1-z)}\Phi_q(z,\zeta,W^2) \right
\}.\nonumber\\
\label{hdtensor2}
\end{eqnarray}
The function $\Phi_q(z,\zeta,W^2)$ is the generalized distribution
amplitude defined as~\cite{Diehl98}:
\begin{eqnarray}
\Phi_q(z,\zeta,W^2)&=&\int \frac{dx^-}{2\pi} e^{iz(P^+
x^-)}\nonumber\\
&&\times\langle
\pi(p_1)\pi(p_2)|\bar{\psi}(x^-)\gamma^+\psi(0)|0\rangle.
\end{eqnarray}
Note that because the final hadrons are two pions (pseudoscalar
mesons), the third and fourth lines of (\ref{hardcoef}) vanishes by
parity invariance of the strong interactions.

Before further discussion it is useful to recall some symmetry
properties of $2\pi$DAs. Charge conjugation invariance of the strong
interactions implies~\cite{Diehl98}:
\begin{equation}
\Phi_q^{\pi\pi}(z,\zeta,W^2)=-\Phi_q^{\pi\pi}(1-z,1-\zeta,W^2).
\end{equation}
Following the notation in~\cite{Diehl00} the $C$-even and $C$-odd
parts (In \cite{Polyakov99} they are also called isoscalar and
isovector parts respectively) of $2\pi$ DA can be defined as
\begin{equation}
\Phi_q^{\pm}(z,\zeta,W^2)=\frac{1}{2}[\Phi_q^{\pi\pi}(z,\zeta,W^2)
\pm \Phi_q^{\pi\pi}(z,1-\zeta,W^2)],\label{cpm}
\end{equation}
where the superscript $\pi \pi$ represents $\pi^0 \pi^0$ or $\pi^+
\pi^-$. Then
\begin{equation}
\Phi_q^{\pi\pi}(z,\zeta,W^2)=\Phi_q^{+}(z,\zeta,W^2)+\Phi_q^{-}(z,\zeta,W^2).\label{cdecomp}
\end{equation}
The properties of $\Phi_q^{\pm}(z,\zeta,W^2)$ under the interchange
$z \rightarrow 1-z$ and $\zeta \rightarrow 1-\zeta$ can be easily
derived from (\ref{cdecomp})~\cite{Polyakov99}:
\begin{eqnarray}
\Phi_q^{+}(z,\zeta,W^2)&=&-\Phi_q^{+}(1-z,\zeta,W^2)\nonumber\\
&=&\Phi_q^{+}(z,1-\zeta,W^2),\label{ceven}\\
\Phi_q^{-}(z,\zeta,W^2)&=&\Phi_q^{-}(1-z,\zeta,W^2)\nonumber\\
&=&-\Phi_q^{-}(z,1-\zeta,W^2).\label{codd}
\end{eqnarray}
The $\zeta \rightarrow 1-\zeta$ exchange means the interchange of
the two pions, thus in the case in which the final two pions are
$\pi^0 \pi^0$ we get
\begin{equation}
\Phi_q^{\pi^0 \pi^0}(z,\zeta,W^2)=\Phi_q^{\pi^0
\pi^0}(z,1-\zeta,W^2),
\end{equation}
Therefore there is only a $C$-even part for $\Phi_q^{\pi^0
\pi^0}(z,\zeta,W^2)$. This is true because the $\pi^0 \pi^0$ state
is a $C$-even state.

Now we turn back to Eq.~(\ref{hdtensor2}). The hard coefficient in
the first term inside the brackets is antisymmetric under $z
\rightarrow 1-z$, while the second one is symmetric under $z
\rightarrow 1-z$. Then by virtue of
(\ref{cdecomp},\ref{ceven},\ref{codd}), we have
\begin{eqnarray}
\hspace{-0.5cm}&&T^{\mu \nu}=-\frac{g}{4 \cos\theta_W}\sum_q ee_q
\left \{c^q_V g_\perp^{\mu \nu} \int_{0}^{1}
dz\frac{2z-1}{z(1-z)}\right
.\nonumber\\
& & \times\Phi_q^+(z,\zeta,W^2)\left . -ic^q_A\epsilon_\perp^{\mu
\nu} \int_{0}^{1} dz\frac{1}{z(1-z)}\Phi_q^-(z,\zeta,W^2) \right
\}.\nonumber\\
\label{hdtensor3}
\end{eqnarray}

The first term in (\ref{hdtensor3}) probes the $C$-even part of the
$2\pi$DAs, which is the same as that in the two-photon process
except for the different coupling. Here the main difference with the
two-photon process, which can only project out the $C$-even part of
the $2\pi$DAs, is an additional term, the second term in
(\ref{hdtensor3}), which can probe the $C$-odd part of $2\pi$DAs.

The amplitude of $Z\rightarrow \pi\pi\gamma$ can be calculated from
\begin{equation}
\mathcal{A}_{i,j}^{Z\rightarrow
\pi\pi\gamma}=\epsilon^{(i)}_\alpha\epsilon^{\prime (j)*}_\beta
T^{\alpha\beta}(z,\zeta,W^2),
\end{equation}
where $\epsilon$ and $\epsilon^\prime$ are the polarization vectors
of $Z$ boson and real photon, respectively. In our frame these
vectors have the form:
\begin{eqnarray}
\epsilon^{(\pm)}_\mu&=&\left(0,\frac{\mp
1}{\sqrt{2}},\frac{-i}{\sqrt{2}},0\right
),\epsilon_\mu^0=\left(\frac{|\Vec q|}{Q},0,0,\frac{q_0}{Q}\right
)\nonumber\\ \epsilon^{\prime(\pm)}_\mu&=&\left(0,\frac{\mp
1}{\sqrt{2}},\frac{-i}{\sqrt{2}},0\right ).
\end{eqnarray}
Similar to the two-photon process, in leading twist only the
transverse polarization contributes, and the polarizations of the
$Z$ boson and the photon must be the same. This follows from angular
momentum conservation in the collinear approximation.

\section{Cross-section and numerical result}

The pion pair production at the $Z$-pole can be realized in the
$e^+e^-$ annihilation process $e^+e^-\rightarrow Z\rightarrow \pi
\pi \gamma$. It is interesting to mention that except for the fact
that the process occurs at the $Z$-pole, this process is related by
crossing to the virtual Compton process $e \pi\rightarrow
e\pi\gamma$,  as well as the process $e\gamma\rightarrow e\pi\pi$.

With the polarizations of the final photon summed the amplitude of
the process is calculated from
\begin{eqnarray}
&&\mathcal{A}_{e^+e^-\rightarrow Z\rightarrow
\pi\pi\gamma}=\frac{g}{2\cos\theta_W}\sum_{i,j}\bar{v}(l^\prime)V^\mu
u(l) \epsilon^{(i)*}_\mu\nonumber\\
&&\hspace{1cm}\times\frac{1}{q^2-M_Z^2-i\Gamma_Z
M_Z}\mathcal{A}_{i,j}^{Z\rightarrow \pi\pi\gamma}.
\end{eqnarray}
Here $V^\mu=\gamma^\mu(c^l_V-\gamma_5c^l_A)$ is the $Z$-boson-lepton
vertex, and $c^l_V$ and $c^l_A$ have the same form as shown in
(\ref{cqv}), and (\ref{cqa}) where $T_3^l$ for the electron is
$-1/2$.

The differential cross-section for the process $e^+e^-\rightarrow
Z\rightarrow\pi\pi\gamma$ is expressed as
\begin{eqnarray}
d\sigma_{e^+e^-\rightarrow Z\rightarrow\pi\pi
\gamma}&=&\frac{1}{2S_{ee}}\frac{d^3p_1}{(2\pi)^32p_1^0}
\frac{d^3p_2}{(2\pi)^32p_2^0}\frac{d^3q^\prime}{(2\pi)^32q^{\prime
0}}\nonumber\\
& &\hspace{-1.5cm}\times|\mathcal{A}_{e^+e^-\rightarrow Z\rightarrow
\pi\pi \gamma}|^2\delta^4(q-p_1-p_2-q^\prime),
\end{eqnarray}
where $s_{ee}=(l+l^\prime)^2=Q^2$ is the center of mass energy
squared of the lepton pair. We then rearrange the kinematics of the
phase space to~\cite{PDG04}
\begin{eqnarray}
d\sigma_{e^+e^-\rightarrow
Z\rightarrow\pi\pi\gamma}&=&\frac{1}{16(2\pi)^5S_{ee}Q}|\mathcal{A}_{e^+e^-\rightarrow
Z\rightarrow \pi\pi\gamma}|^2\nonumber\\
&&\times|\Vec p^*||\Vec q^\prime|dWd\Omega^*d\Omega^\prime,
\end{eqnarray}
in which ($|\Vec p^*|,\Omega^*$) is the momentum of pion 1 in the
c.m. frame of the pion pair, and $\Omega^\prime$ is the angle of the
photon in the rest frame of the $Z$ boson (that is, the c.m. frame
of $e^+e^-$).  $|\Vec p^*|$ and $|\Vec q^\prime|$ are given by
\begin{eqnarray}
|\Vec p_1^*|&=&\frac{1}{2}\sqrt{W^2-4m_{\pi}^2}=\frac{\beta W}{2},\\
|\Vec q^\prime|&=&\frac{Q^2-W^2}{2Q}.
\end{eqnarray}

The amplitude square is
\begin{eqnarray}
\hspace{-0.5cm}&&|\mathcal{A}_{e^+e^-\rightarrow Z\rightarrow
 \pi\pi\gamma}|^2=\frac{e^2a_w^2Q^2}{(Q^2-M_Z^2)^2+\Gamma_Z^2 M_Z^2}
 \nonumber\\
 &&\times \left \{(c_V^{l2}+c_A^{l2})I_1(y)\left [|V(\cos\theta,W^2)|^2 +
|A(\cos\theta,W^2)|^2\right]\right.\nonumber\\
&&\left.+4c_V^lc_A^lI_2(y) \textrm{Im}\{V^*(\cos\theta,W^2)
A(\cos\theta,W^2)\}\right \}.
\end{eqnarray}
The symbols in the above equation are
\begin{eqnarray}
a_w&=&\frac{e^2}{4\sin^2\theta_W\cos^2\theta_W},\\
I_1(y)&=&(\frac{1}{2}-y+y^2),\\
I_2(y)&=&(1-2y),\\
V(\cos\theta,W^2)&=&\sum_q e_q c_V^q \int_{0}^{1}
dz\frac{2z-1}{z(1-z)}\nonumber\\
&&\times\Phi_q^+(z,\zeta(\cos\theta),W^2),\\
A(\cos\theta,W^2)&=&\sum_q e_q c_A^q  \int_{0}^{1}
dz\frac{1}{z(1-z)}\nonumber\\
&&\times\Phi_q^-(z,\zeta(\cos\theta),W^2),
\end{eqnarray}
 where $y=l\cdot q^\prime/q\cdot
q^\prime$. In the c.m. frame of the lepton pair (see
Fig.~\ref{kine}b) there is the relation $y=(1+\cos\theta^\prime)/2$,
where $\theta^\prime$ is the angle of the photon with respect to the
momentum of the incoming positron. Thus $y$ is in the interval
$0<y<1$.

\begin{figure}

\begin{center}
\scalebox{0.8}{\hspace{-0.5cm}\includegraphics{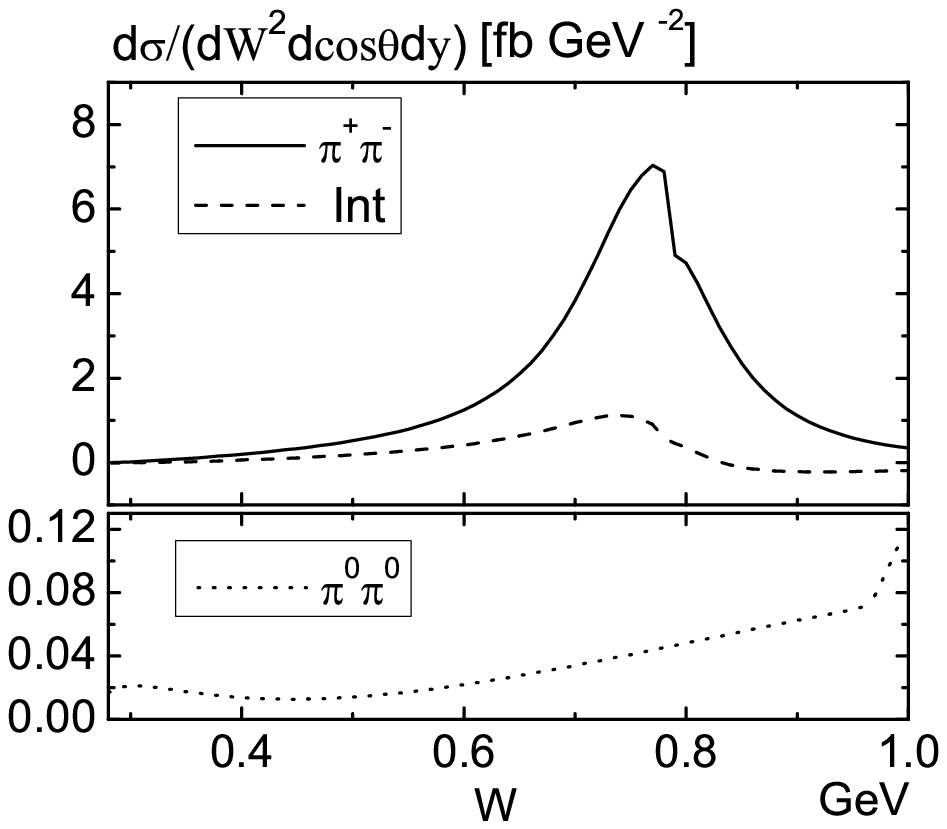}}
\caption{\small The $W$ dependence of the differential cross-section
for charged pion pair (solid line in upper panel) and neutral pion
pair (dotted line in lower panel) production at the $Z$-pole at
$\theta=10^\circ$ and $y=0.1$. In the upper panel we also show the
contribution from the interference (dashed line) between the $C$-odd
and $C$-even parts of the $2\pi DA$. } \label{wdepend}
\end{center}

\end{figure}

\begin{figure}

\begin{center}
\scalebox{0.8}{\hspace{-0.5cm}\includegraphics{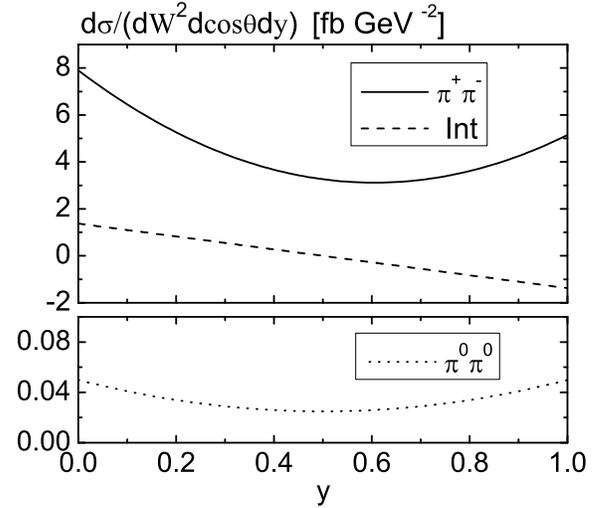}}
\caption{\small The $y$ dependence of the differential cross-section
for charged pion pair (solid line in upper panel) and neutral pion
pair (dotted line in lower panel) production at the $Z$-pole at
$W=0.75$ GeV and $\theta=10^\circ$. In the upper panel we also show
the contribution from the interference (dashed line) between the
$C$-odd and $C$-even parts of the $2\pi DA$.} \label{ydepend}
\end{center}

\end{figure}

Then we write the differential cross-section as
\begin{eqnarray}
&&\hspace{-0.5cm}\frac{d\sigma^{e^+e^-\rightarrow Z \rightarrow
\pi\pi\gamma}}{dW^2 d\cos \theta d\phi
dy}=\frac{e^2a_w^2\beta(Q^2-W^2)}{128(2\pi)^4((Q^2-M_Z^2)^2+\Gamma_Z^2
M_Z^2)Q^2}\nonumber\\
&&\times\left
 \{I_1(y)(c_V^{l2}+c_A^{l2})\left  [|V(\cos\theta,W^2)|^2 +
|A(\cos\theta,W^2)|^2\right]\right .\nonumber\\
&&\left .+4I_2(y)c_V^lc_A^l \textrm{Im}\{V^*(\cos\theta,W^2)
A(\cos\theta,W^2)\}\right \}.\label{dcs}
\end{eqnarray}
The three terms in the above equation give the contribution from the
$C$-even channel, the $C$-odd channel and the interference of the
two channels, respectively.

\begin{figure}

\begin{center}
\scalebox{0.8}{\hspace{-0.5cm}\includegraphics{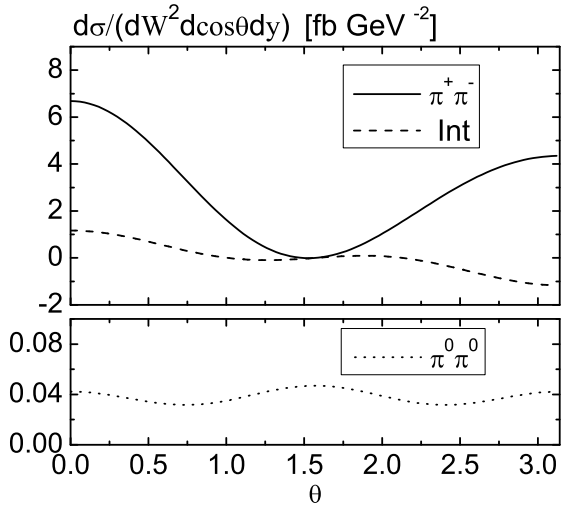}}
\caption{\small The $\theta$ dependence of the differential
cross-section for charged pion pair (solid line in upper panel) and
neutral pion pair (dotted line in lower panel) production at the
$Z$-pole at  $W=0.75$ GeV and $y=0.1$. In the upper panel we also
show the contribution from the interference (dashed line) between
the $C$-odd and $C$-even parts of the $2\pi DA$. }
\label{thetadepend}
\end{center}

\end{figure}

In the case of the charged pion pair production, however, there is
also contribution from a Bremsstrahlung type process that
$e^-e^+\rightarrow \gamma^*\gamma\rightarrow \pi^+\pi^-\gamma$,
where a time-like photon produced in a QED process decays into the
pion pair. The differential cross-section of this process in the
kinematics that we consider ($Q^2=M_Z^2 \gg W^2$) has the form
\begin{eqnarray}
\frac{d\sigma^B}{dW^2 d\cos \theta d\phi dy}
        &=& \frac{ e^6 \beta (Q^2 - W^2) }{ 128 (2 \pi)^4 Q^4 W^2}\nonumber\\
        &&\hspace{-2cm}\times \frac{ 2 \beta^2 I_1 (y)}{y(1-y)}|F_\pi(W^2)|^2 \sin^2 \theta.\label{bremcs}
\end{eqnarray}
To complete the analysis, we give the the interference term between
the production at the $Z$-pole and the Bremsstrahlung process:
\begin{eqnarray}
\frac{d\sigma^I}{dW^2 d\cos \theta d\phi dy}
        &=&\frac{e^4a_w\beta(Q^2-W^2)}{128(2\pi)^4Q^4\Gamma_Z W}\nonumber\\
        &&\hspace{-3.5cm}\times\left [\frac{2\beta c_v^l I_1(y)}{\sqrt{y(1-y)}}\textrm{Re}\{F_\pi(W^2)^*V\}\sin\theta
         \cos\phi\right . \nonumber\\
         &&\hspace{-3.5cm}-\frac{2\beta c_a^l I_1(y)}{\sqrt{y(1-y)}}\textrm{Re}\{F_\pi(W^2)^*A\}\sin\theta
         \cos\phi\nonumber\\
         &&\hspace{-3.5cm}-\frac{\beta c_v^l I_2(y)}{\sqrt{y(1-y)}}\frac{Q^2}{Q^2-W^2}\textrm{Im}\{F_\pi(W^2)^*A\}\sin\theta
         \sin\phi\nonumber\\
         &&\hspace{-3.5cm}\left.+\frac{2\beta c_a^l  I_1(y)}{\sqrt{y(1-y)}}\textrm{Im}\{F_\pi(W^2)^*V\}\sin\theta
         \sin\phi\right ].\label{intecs}
\end{eqnarray}
In Eq.~(\ref{bremcs}) and (\ref{intecs}) the $W/Q$ suppressed terms
are not kept, since they are irrelevant at the $Z$-pole energy
scale.

We now use a simple model for the 2$\pi$DAs to estimate the
cross-section of pion pair production at the $Z$-pole through
Eq.~(\ref{dcs}). In the following we use $\Phi_q^{\pm}(z,\zeta,W^2)$
to represent the $C$-even/odd part of
$\Phi_q^{\pi^+\pi^-}(z,\zeta,W^2)$. Isospin invariance
implies~\cite{Diehl00}
\begin{eqnarray}
\Phi_q^{\pi^0\pi^0}(z,\zeta,W^2)=\Phi_q^{+}(z,\zeta,W^2).
\end{eqnarray}
Therefore, only the first term in (\ref{dcs}) contributes to the
neutral pion pair production. Here we only consider the
contributions from the $u$ and $d$ quark. Another consequence of
isospin invariance is
\begin{eqnarray}
\Phi^{+}_u=\Phi^{+}_d,~~~~\Phi^{-}_u=-\Phi^{-}_d,
\end{eqnarray}
Diehl et.al. gave a simple model for $\Phi^{+}_q$ based on the
Legendre polynomials expansion of the GDAs, as
follows~\cite{Diehl00}
\begin{eqnarray}
\Phi^{+}_u(z,\zeta(\cos\theta),W^2)&=&10z(1-z)(2z-1)R_\pi\nonumber\\
&&\hspace{-3cm}\times\left [ -\frac{3-\beta^2}{2} e^{i\delta_0(W^2)}
+ \beta^2e^{i\delta_2(W^2)} P_2(\cos\theta)\right ], \label{phiplus}
\end{eqnarray}
in which $\delta_0(W^2)$ and $\delta_2(W^2)$ are the S wave and D
wave phase shifts of elastic $\pi \pi$ scattering, respectively. The
analysis for these phase shifts is available, for example, in
Ref.~\cite{em74}. This model is valid in the regime $W<1$ GeV, and
we restrict our calculation to this regime. For $\Phi^{-}_q$ we use
the asymptotic form at large $Q^2$, corresponding to the P wave
contribution~\cite{Polyakov99}:
\begin{equation}
\Phi^{-}_u(z,\zeta(\cos\theta),W^2)=6z(1-z)\beta
P_1(\cos\theta)F_\pi(W^2),\label{phiminus}
\end{equation}
where $F_\pi(W^2)$ is the time-like pion form factor, for which we
use the parametrization $N=1$ given in Ref.~\cite{ks90}.

\begin{figure*}

\begin{center}
\scalebox{1.08}{\hspace{-0.5cm}\includegraphics{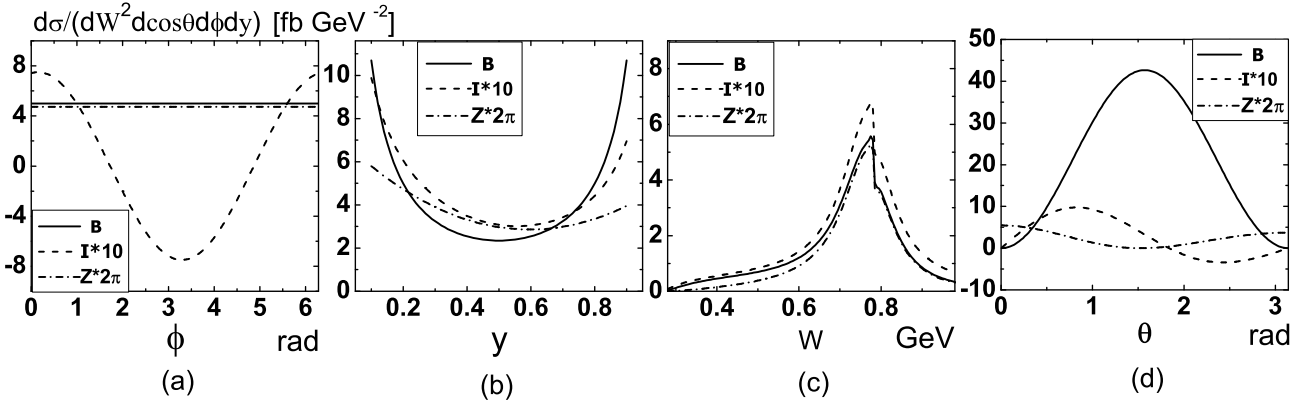}}
\caption{\small The differential cross-sections contributed by the
Bremsstrahlung process (solid line) and its interference (dashed
line) with the production at the $Z$-pole at $Q=M_Z$. The dashed
dotted line shows the cross-section contributed by the production at
the $Z$-pole scaled by a factor $2\pi$ for comparison. (a) the
$\phi$ dependence of the cross-section at $y=0.2$,
$\theta=20^\circ$,$W=0.75\,\textrm{GeV}$; (b) the $y$ dependence of
the cross-section at $\phi=45^\circ$, $\theta=20^\circ$ and
$W=0.75\,\textrm{GeV}$; (c) The $W$ dependence of of the
cross-section at $\phi=45^\circ$, $y=0.2$ and $\theta=20^\circ$; (d)
the $\theta$ of the cross-section at $\phi=45^\circ$,
 $y=0.2$ and $W=0.75\,\textrm{GeV}$. The interference contribution
 has been scaled by a factor 10.} \label{bremin}
\end{center}

\end{figure*}

\begin{figure}

\begin{center}
\scalebox{0.9}{\hspace{-0.5cm}\includegraphics{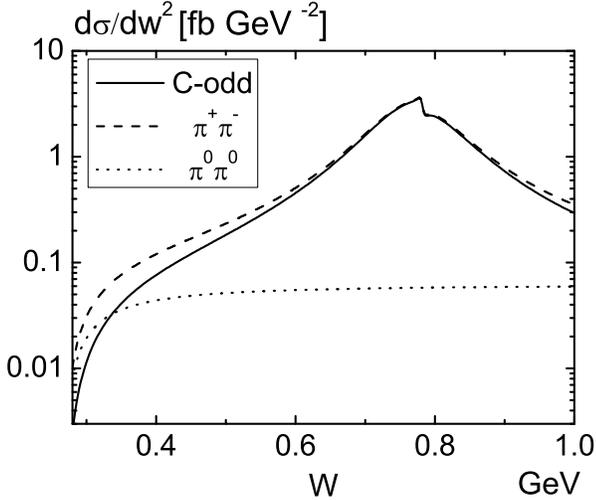}}
\caption{\small The $W$ dependence of the contribution from
$\Phi^-_q$ (solid line) to the charge pion pair production at the
$Z$-pole, after integrating over $y$. The dashed and dotted lines
show the neutral and charged pion pair production in the same case,
respectively.} \label{cs}
\end{center}

\end{figure}

In Fig.~\ref{wdepend} we show the $W$ dependence of the differential
cross-section for the neutral pion and charged pion pair production
using the 2$\pi$DAs given in (\ref{phiplus}) and (\ref{phiminus}).
We also show the contribution from the interference term of the
$C$-odd and $C$-even parts of the $2\pi DA$ to the charged pion
production at the $Z$-pole. The corresponding $y$ dependence and
$\theta$ dependence of the differential cross-section are given in
Fig.~\ref{ydepend} and Fig.~\ref{thetadepend}, respectively. The
figures show that the cross-section of the charged pion pair
production is much larger than that of the neutral pion pair, nearly
$1 \sim 2$ order of magnitude in the region $0.4$ GeV $<W<0.9$ GeV.
The contribution of the interference term is significant compared
with the cross-section of the neutral pion pair production, but
still several times smaller than that of the charged pion
production. One can conclude that most of the contribution to the
charged pion production comes from
 $\Phi^-_q$. This is expected from the large magnitude
of the time-like form factor of the pion and the different $z$
dependence of the $\Phi^{\pm}_q$ given in (\ref{phiplus}) and
(\ref{phiminus}).

In the $W$ range which Fig.~\ref{wdepend} covers, the contributions
from the Bremsstrahlung process and its interference with the
production at the $Z$-pole are not suppressed, due to the factor
$W^2$ and $W$ in the denominator of Eq.~(\ref{bremcs}) and
Eq.~(\ref{intecs}). The differential cross-sections contributed by
these two terms at $Q=M_Z$ are depicted in Fig.~\ref{bremin}, where
the interference contribution has been scaled by a factor 10. In
Fig.~\ref{bremin} we also give a comparison between Bremsstrahlung
with the production at the $Z$-pole which is scaled by a factor
$2\pi$. The numerical results show that the contribution from the
Bremsstrahlung process is several times larger than that from the
production at the $Z$-pole (note that in Fig.~\ref{wdepend},
\ref{ydepend} and \ref{thetadepend} the azimuthal angle $\phi$ has
been integrated over).  The interference term given in
(\ref{intecs}) is about one order of magnitude less than the
Bremsstrahlung process.

The large contribution of the $C$-odd channel production at the
$Z$-pole compared to the production at the $C$-even channel provides
an opportunity to obtain information on
 $\Phi^-_q$. The same part also contributes in
hard electroproduction~\cite{Polyakov99,lppsk} and charm or $B$
meson decay~\cite{cl04}. However, in these cases the GDAs convolute
with non-perturbative GPDs or meson distribution amplitudes. In the
pion pair production case at the $Z$-pole since there are no
hadronic states in the initial state, GDAs only convolute with the
perturbative calculable hard coefficients, thus $\Phi^-_q$ can be
accessed more cleanly in this case.

We now discuss how to obtain the contribution coming from
$\Phi^{-}_q$ in $e^+e^-$ annihilation process at the $Z$-pole. First
one confront the large contribution from the Bremsstrahlung process
which need to be carefully treated. Because the modulus of $F_\pi$
has been well measured, the contribution from Bremsstrahlung can be
predicted quite well by Eq.~(\ref{bremcs}). The interference between
the Bremsstrahlung process and the production at the $Z$-pole is
eliminated by integrating over azimuthal angle $\phi$. Therefore the
$Z$-pole contribution can be separated out. Again notice that the
interference term in (\ref{dcs}) vanishes after integrating over $y$
(or alternatively, integrating over $\theta$), therefore the
difference of the differential cross-sections of $\pi^+\pi^-$ and
$\pi^0\pi^0$ production at the $Z$-pole (after integrating over $y$)
is
\begin{eqnarray}
&&\frac{d\sigma^{e^+e^-\rightarrow
Z\rightarrow{\pi^+\pi^-\gamma}}}{dW^2 d\cos \theta
}-\frac{d\sigma^{e^+e^-\rightarrow
Z\rightarrow\pi^0\pi^0{\gamma}}}{dW^2 d\cos \theta }\nonumber\\
&=&\frac{e^2a_w^2(Q^2-W^2) (c_V^{l2}+c_A^{l2})
|A(\cos\theta,W^2)|^2}{384(2\pi)^3((Q^2-M_Z^2)^2+\Gamma_Z^2
M_Z^2)Q^2}.\label{coddm}
\end{eqnarray}
The above result depends only on the moment of $\Phi^{-}_q$. Thus
combining the measurement of charged pair production and neutral
pair production at the $Z$-pole, the moment of $\Phi^{-}_q$ is
obtained. Fig.~\ref{cs} shows the contribution of $\Phi^{-}_q$ to
the charged pair production which is calculated from
Eq.~(\ref{coddm}). Due to the relative small cross-section, the
extraction of $\Phi^{-}_q$ is difficult through the pion pair
production at the $Z$-pole based on the LEP-I data~\cite{lep}, while
it is feasible at the possible Giga-$Z$ option~\cite{ilc} of the
planned International Linear Collider, at which the integrated
luminosity can reach $100\,\textrm{fb}^{-1}$ in about one yr of
running.

\section{Summary}

We have shown that pion pairs with small invariance mass, produced
at the $Z$-pole through $e^+e^- \rightarrow Z\rightarrow
\pi\pi\gamma$, can be factorized into a hard scattering coefficient
convoluted with the non-perturbative two-pion distribution
amplitudes. We calculated the cross-sections for the production of a
charged pion pair and of a neutral pion pair in leading order. The
comparison of these two cases show that in the regime 0.4 GeV $< W<
0.9$ GeV the cross-section of the charged pair production is much
larger than that of neutral pair production, showing that the
process is dominated by the contribution from the $C$-odd part of
the $2\pi$DAs. By virtue of isospin symmetry, one can combine the
measurements of the charged pair and neutral pair production at the
$Z$-pole to access the $C$-odd part of the $2\pi$DAs cleanly. The
pion pair production at the $Z$-pole at $e^+e^-$ facilities can open
opportunities to obtain valuable information on the 2$\pi$DAs,
especially the $C$-odd part of 2$\pi$DAs, and can be viewed as the
useful complement to the two-photon process.

\begin{acknowledgments}
This work is supported by Fondecyt (Chile) under Projects
No.~3050047 and No.~1030355.
\end{acknowledgments}

\end{document}